\documentclass[3p,times]{elsarticle}

\usepackage{amsmath}
\usepackage{amssymb}
\usepackage{accents}
\usepackage{bm}
\usepackage{graphicx}
\usepackage{color}
\usepackage{dsfont}
\usepackage{array}
\usepackage{natbib}
\usepackage{geometry}
\usepackage{amsfonts}
\usepackage{longtable}
\usepackage{lscape}
\usepackage[table]{xcolor}
\usepackage{subcaption}

\newcommand{\dd}{{\rm d}}

\usepackage{amssymb}
\usepackage[figuresright]{rotating}

\begin{document}

\begin{frontmatter}




\title{Primordial magnetogenesis before recombination}


\author{Oph\'elia Fabre}\ead{ophelia.fabre@iisertvm.ac.in}
\author{S. Shankaranarayanan}\ead{shanki@iisertvm.ac.in}

\address{School of Physics, Indian Institute for Science Education and Research Thiruvananthapuram (IISER-TVM), Trivandrum 695016, India}

\begin{abstract}
The origin of large magnetic fields in the Universe remains currently unknown. We investigate here a mechanism before recombination based on known physics. The source of the vorticity is due to the changes in the photon distribution function caused by the fluctuations in the background photons. We show that the magnetic field generated in the MHD limit, due to the Coulomb scattering, is of the order $10^{-49}$~G on a coherence scale of 10 kpc. We explicitly show that the magnetic fields generated from this process are sustainable and are not erased by resistive diffusion. We compare the results with current observations and discuss the implications. Our seed magnetic fields are generated on small scales whereas the main mechanisms studied in the literature are on scale bigger than 1 Mpc. However, compared to more exotic theories generating seed magnetic fields on similar scales, the strength of our fields are generally smaller.
\end{abstract}


\end{frontmatter}

% \linenumbers

\section{Introduction}\label{introduction}

The origin of galactic and inter-galactic large coherence-scale
magnetic fields remains largely unknown \cite{widrow2002origin,kronberg1994extragalactic,beck1996galactic}. 
Galactic fields of micro-Gauss  
strength and coherence length between 1~kpc and 10~kpc, and inter-galactic 
magnetic fields of strength $10^{-7}-10^{-6}$~G and coherence length 
between 10~kpc and 1~Mpc, have been observed \cite{widrow2002origin,kronberg1994extragalactic,beck1996galactic,enqvist1998primordial,grasso2001magnetic,battaner2000physics,dolgov2002generation,
  dolgov2003magnetic,giovannini2004magnetized}. While the above magnetic 
  field measurements are upper bounds, FERMI measurement of
gamma-rays emitted by blazars seem to provide lower bound of the 
order of $10^{-15}~\rm{G}$ in voids \cite{neronov2010evidence,dermer2011time}.

Several mechanisms have been proposed to explain the origin of the
magnetic fields on large scales. These models can broadly be
categorized into early times and late times. In the case of late time
models, fields generated in the proto-galaxies are spilled to the
inter-galactic medium
\cite{lesch1994protogalactic,chiba1994galactic,dolag2005constrained},
while in the case of an early time models
\cite{grasso2001magnetic,shaposhnikov2005primordial}, the fields are
generated during inflation
\cite{turner1988inflation,ratra1992cosmological,garretson1992primordial,dolgov1993breaking,dolgov1993electric},
GUT/EW phase transition
\cite{vachaspati1991magnetic,bradenberger1992superconducting,enqvist1993primordial},
radiation domination or post-recombination era
\cite{harrison1973magnetic,baierlein1978amplification,berezhiani2004generation}. For
example, magnetic fields generated from density perturbation have been
investigated
\cite{fenu2011seed,takahashi2005magnetic,matarrese2005large}, as well
as stochastic background of magnetic fields
\cite{banerjee2004evolution,paoletti2011cmb,ade2013planck,planckplanck,seshadri2009cosmic}
and inhomogeneous magnetic fields
\cite{seshadri2001cosmic,subramanian1998microwave}. Both these
 --- early and late times --- categories have advantages and
disadvantages. For instance, the early time models can not generate
required magnetic field strength, while the late time models can not
generate fields with the required coherence length. All the models
provide the seed fields which need to be amplified by the dynamo
mechanism to maintain the observed galactic and inter-galactic
magnetic field strength.
 
In this work, we focus on the generation of large-scale seed magnetic
fields before recombination. The model
we consider is based on the work of \cite{berezhiani2004generation}
and does not involve any new physics. The source of vorticity is due
to the changes in the photon distribution function caused by the
fluctuations in the background photons. The model studied in
\cite{berezhiani2004generation} suffers from several problems. First,
the vorticity was under-estimated: the vorticity is a second-order
physical quantity and some of the second-order contributions were not taken
into account. We correct this by expanding the Boltzmann equation up to 
second order. Second, the electrical conductivity during recombination 
was taken to be arising from Thomson scattering whereas we demonstrate that in the temperature range $20~{\rm eV}<T<100~{\rm eV}$,
the conductivity linked to the Coulomb interaction dominates the
magneto-hydrodynamics equation that generates magnetic fields (see
Eq.~(\ref{MHD})). However, this effect was corrected in \cite{berezhiani2013dark}. Third, and crucial point, the current inside the plasma was over-estimated. Finally, after recombination, the conductivity of
the Universe is very high. The magnetic flux can thus be considered as
frozen (\cite{martin2008generation}) and the strength of the magnetic
fields generated before recombination decrease until the collapse of
the first structures. This effect was not taken into account in Ref.~\cite{berezhiani2004generation}. We include this effect before estimating a dynamo amplification by proto-galaxies
\cite{lesch1994protogalactic} and galaxies.

In Section~(\ref{section2}), we present in detail the model of generation of 
seed magnetic field before recombination. In Section~(\ref{parameters}), we compute the main parameters of our model, like the vorticity (\ref{ss_vorticity}) and the conductivity (\ref{ss_conductivity}), to obtain in Section~(\ref{resulting_B}) a theoretical estimation of the strength and the coherence length of the magnetic field seeds generated by our model. In Section~(\ref{comparison}), we compare and contrast our model with other models in the literature. Finally, in Section~(\ref{conclusion}), we discuss the importance of our work in the light of the recent measurement. To keep the continuity, we have included most of the calculations in Appendices.

\section{The model}\label{section2}

We focus on the pre-recombination era in the temperature range from 
$100$~eV to $20$~eV. It is important to note that $E=13.6$~eV is the
ionization energy of the hydrogen, hence below this energy the
fraction of free electron $x_e$ will start decreasing slowly until the
beginning of recombination (around $0.25$~eV) when $x_e$ becomes tiny. 
For the energy range 20~eV$<T<$100~eV,
Universe is well described by a hot and dense plasma at equilibrium composed
of protons, electrons, photons and dark matter. We neglect ionized
helium and any gravitational interaction. We also assume that dark
matter does not interact with the plasma. Because of the global
electrical neutrality of the medium, the number density of electrons
($n_e$) and the number density of protons ($n_p$) are equal at $0^{\rm th}$ order. Let $i\in\{e,p\}$, where $e$ stands for electrons and $p$ for protons. We expand the number density of particles and their velocity up to second order as in \cite{takahashi2008electromagnetic}, so that
\begin{eqnarray}
n_i&=&n+n^{(1)}_i+n^{(2)}_i +o(n_i^{(3)}) \\
\vec{v}_j&=&\vec{v}+{\vec{v}_j}^{(1)}+\vec{v}^{(2)}_j +o(v_j^{(3)}) \\
\delta\vec{v}_{ij}&=&\vec{v}_i-\vec{v}_j \\
\delta n_{ij}&=& n_i -n_j.
\end{eqnarray}
%
%These two quantities are proportional to $T^3$ (cf. Fermi-Dirac distribution). 
We define $\beta=n_e/n_\gamma$ where $n_\gamma$ is 
the number density of photons. CMB and Big Bang Nucleo-synthesis 
(BBN) constraint the value of $\beta \simeq 6\times
10^{-10}$ \cite{kaplinghat2000precision,cyburt2003primordial,ade2013planck}. Although electrons and protons are freely moving in the plasma, the plasma itself has to be electrically neutral on the whole.

Unlike Ref.~\cite{berezhiani2004generation}, we consider two
scattering processes involving the charged particles. We include 
Thomson scattering, which depicts the
diffusion of photons on free electrons, and Coulomb
scattering of free electrons on protons. Let $V_p$,
$V_{e^-}$ and $V_\gamma$ denote the velocities of protons, 
electrons and photons, respectively. Due to the respective inertia of each of these
particles, we could expect, like in \cite{berezhiani2004generation}, that $V_p \ll V_{e^-} \ll V_\gamma$, approximate that the protons do not move compared to
the electrons, and that the induced current, and the seed magnetic
fields $\vec{B}_{\rm seed}$, are generated by the differential
motion of electrons inside the plasma, with the current being thus given by $\vec{J}\approx-en_e \vec{v}_e$. But this approach is not correct, because thanks to the tight-coupling approximation and the efficiency of Coulomb scattering on the temperature range considered, we will have $V_p \lessapprox V_{e^-}\lessapprox V_\gamma$. Because of the inaccurate estimation of the current, $J$ and the seed magnetic fields generated were over-estimated in \cite{berezhiani2004generation}. The correct expression for the current in the plasma will be 

\begin{eqnarray}
\vec{J} &=& e\left(n_p \vec{v}_p-n_e \vec{v}_e\right)\\
        &\approx& e\left(n \delta\vec{v}_{pe}+\delta n_{pe} \vec{v} \right).
\end{eqnarray}

Finally, as the photons are 
the most mobile particles in the plasma, the bulk velocity $v$ of the plasma 
can be approximated by $v_\gamma\propto\int \dd^3\vec{p}~V_\gamma f_\gamma$, 
where $f_\gamma$ is the distribution function of photons.

%Due to the efficient scattering of photons on electrons, a uniform distribution 
%of photons do not generate vorticity. When the primordial scalar perturbations 
%\marginpar{move to Sec. (IIIB)}
%re-enter the Hubble radius, they produce changes in the photon distribution. As a
%consequence, the electrons are scattered non-uniformly resulting in the generation
%of small vorticity $\Omega$.

\section{Generation of primordial magnetic field}\label{parameters}

From \cite{takahashi2008electromagnetic} and \ref{plasma_equations}, we have the following equation describing magnetic fields in the plasma
\begin{eqnarray}
\partial_t \vec{B}\approx\vec{\nabla}\times\left(\vec{v}_e \times \vec{B}\right)&+&\frac{1}{e}\vec{\nabla}\times\left(\frac{\vec{\nabla} P_e}{n_e}\right)-\vec{\nabla}\times\frac{m_e}{\tau_{ep}e^2}\left[\frac{\vec{J}}{n_e}\right]-\vec{\nabla}\times\frac{Rm_e}{e\tau_{\gamma e}}\left(\delta\vec{v}_{\gamma e}-\frac{1}{4}\vec{v}_e.\Pi_\gamma\right),
\end{eqnarray}
where $P_e$ is the pressure of the electron fluid, $m_e$ the mass of the electron, $R=\rho_\gamma/n_e m_e$, $\delta \vec{v}_{\gamma e}=\vec{v}_\gamma-\vec{v}_e$, $\tau_{ep}$ (resp. $\tau_{\gamma e}$) the typical time between two collisions of an electron (resp. photons) on protons (resp. electrons); see \ref{ss_conductivity} for more details. $\Pi_{\gamma}$ is the tensor of anisotropic pressure of the photons. As a consequence, we can distinguish two ways of generating the magnetic fields, with different source term. One approach is using curl of the current of the plasma, and thus its vorticity, as a source of magnetic field (see for example \cite{berezhiani2004generation}) in a fluid approximation, \textit{i.e.} in the magneto-hydrodynamics framework by taking into account Coulomb scattering only, so that 
\begin{eqnarray}\label{MHD}
\partial_t \vec{B}_{\rm seed}=\vec{\nabla}\times(\vec{v}\times\vec{B}_{\rm seed})+\frac{1}{\kappa_{\rm Coul}}\vec{\nabla}\times\vec{J},
\end{eqnarray}
where $\kappa_{\rm Coul}=\frac{\tau_{ep}e^2 n_e}{m_e}$ is the conductivity of the medium due to the Coulomb interaction (cf. Section~\ref{ss_conductivity} and \ref{appendix_conductivity}) and $\vec{v}\approx\vec{v}_e$ thanks to the tight-coupling and fluid approximations. The second way to generate magnetic fields is by considering the contribution of the pressure of the electron fluid, Thomson scattering and the tensor of anisotropic pressure (see for example \cite{gopal2005generation}).
\begin{eqnarray}\label{gopal-sethi}
\partial_t \vec{B}_{\rm seed}&=&\vec{\nabla}\times(\vec{v}_e\times\vec{B}_{\rm seed})+\frac{1}{e}\vec{\nabla}\times\left(\frac{\vec{\nabla} P_e}{n_e}\right)-\vec{\nabla}\times\frac{Rm_e}{e\tau_{\gamma e}}\left(\delta\vec{v_{\gamma e}}-\frac{1}{4}\vec{v}_e .\Pi_\gamma\right).
\end{eqnarray}
In this article, we are going to detail the first process and only refer to the second one to compare the two contributions and evaluate if galactic and extra-galactic magnetic fields can be generated by these phenomena.

\subsection{Our model}

Magneto-hydrodynamics (MHD) equation for the evolution of the magnetic
field due to resistive diffusion is given by \cite{1969-Boyd.Sanderson-Book}
\begin{eqnarray}\label{MHD1}
\partial_t \vec{B}_{\rm seed}=\vec{\nabla}\times(\vec{v}\times\vec{B}_{\rm seed})+\frac{1}{\kappa}\vec{\nabla}\times\vec{J},
\end{eqnarray}
where $\kappa$ is the conductivity of the fluid. The first term in the RHS of the above equation is
non-zero only after the generation of the seed field. Hence, the
magnetic field evolves due to the resistive diffusion even in the
absence of plasma flow. Let us take the curl of the current
\begin{eqnarray}
\vec{\nabla}\times\vec{J} &=& e\Big(n \left[\vec{\nabla}\times\delta\vec{v}_{pe}\right]+\left[\vec{\nabla}n\right]\times\delta\vec{v}_{pe}+ \delta n_{pe} \left [\vec{\nabla}\times\vec{v}\right]+ \left[\vec{\nabla}\delta n_{pe}\right]\times\vec{v}\Big).
\end{eqnarray}
To have an idea of the strength of this quantity, we will take into account $J\approx e\delta n_{ep} \left(\vec{\nabla}\times\vec{v}\right)$ only. As a consequence, the magnetic field after a time $t$ is given by
\begin{eqnarray}\label{BMHD}
B_{\rm seed}\approx \int_{t_i}^{t} \dd \tau \frac{e \delta n}{\kappa} \Omega+\int_{t_i}^{t} \dd \tau \frac{n}{\kappa} \left(\Omega_p-\Omega_e\right),
\end{eqnarray}
where $\vec{\Omega}=\vec{\nabla}\times\vec{v}$ is the vorticity of the
plasma, $\vec{\Omega}_j=\vec{\nabla}\times\vec{v}_j$ is the vorticity of the fluid $j$, $t_i$ is the time at which the perturbation enters the Hubble radius and $t$ is the time at
which $B_{\rm seed}$ is generated. We are only going to consider the first integral in Eq.~(\ref{BMHD}). It is
important to note that the first term in the RHS of Eq.~(\ref{MHD}) is
a product of two small quantities and hence the above estimate of the
field is the leading order contribution. We would like to point out some
interesting features regarding Eq.~(\ref{MHD}). First, we have assumed that 
the diffusion time scales, $\tau=\mu_0 \kappa \left(2\pi/\lambda\right)^2$, are large. In other words, if the diffusion time scales 
are comparable to the process time scales, then the generation of the seed magnetic 
fields will not be sustainable and our mechanism will be irrelevant to explain the 
origin of magnetic fields in large scale structures (see \cite{teodoro2008cautionary}).
In \ref{diffusion}, we show that the typical time of diffusion in Eq.~(\ref{MHD}) is larger than the age of the Universe. Second, the amplitude
of the seed field depends on the amplitude of the vorticity generated
due to the processes and inversely proportional to the conductivity of
the processes involved. Third, the processes involved in the
generation of the vorticity need to to be the same processes one takes 
into account in the electrical conductivity. Lastly, Eq.~(\ref{MHD}) implies
that if the vorticity is conserved, the seed magnetic field is also
conserved. 

In the rest of this section, we evaluate different parameters (vorticity, 
conductivity and coherence length) to obtain the primordial magnetic 
field before recombination.
 
\subsection{Vorticity}\label{ss_vorticity}

As discussed above, the cosmological plasma before the recombination is in 
equilibrium. The velocity of the fluid (${\vec v}$) is determined by the 
Boltzmann equation for the photon distribution function $f_\gamma$:
\begin{eqnarray}
\left(\partial_t +\vec{V}_\gamma.\nabla\right)f_\gamma (t,\vec{x},E,\vec{p})
=I_{\rm coll} [f_i],
\end{eqnarray}
where the RHS includes the Thomson scattering between $\gamma$ and $e^{-}$.
The fluid velocity $\vec{v}$ is given by
\begin{eqnarray}
v_k=\frac{1}{\int\dd^3 \vec{p}~f_\gamma^{(0)}}\int\dd^3 \vec{p}~V_k f_\gamma~,
\end{eqnarray}
where $V_k$ is the particle velocity. Since, vorticity $\vec{\Omega}$ is
the curl of the velocity, the second-order terms in temperature
fluctuations can only lead to non-zero vorticity \cite{berezhiani2004generation}. 
To see this, let us expand the photon distribution $f_\gamma=e^{-E/(k_B \, T(\vec{x},t))}$ about the average plasma temperature ($T_0$) i.e., 
\begin{eqnarray}
f_\gamma\approx f^{(0)}_\gamma\Bigg[1&+&\frac{E}{k_B\,T_0}\frac{\delta T}{T_0}
+\Bigg(\frac{E^2}{2(k_B \, T_0)^2}-\frac{E}{k_B \, T_0}\Bigg)\frac{\delta T^2}{T^2_0}+ o\left(\frac{\delta T^3}{T_0^3}\right)\Bigg],
\end{eqnarray}
where $f^{(0)}_\gamma=e^{-E/(k_B T_0)}$. Only the term quadratic in $\delta T/T_0$ can 
lead to non-zero vorticity. For detailed calculation, see \ref{appendix_vorticity}.

For the temperature range $[100\, {\rm eV}, 20 \, {\rm eV}]$, vorticity 
generated due to Thomson scattering is at least 15 orders of magnitude larger 
than the vorticity due to Coulomb scattering. For more details, see \ref{vorticities}. Taking into account all second-order terms, we obtain: 
\begin{eqnarray}\label{omega}
\Omega\approx12\times 
10^3\times c\frac{\ell_\gamma^3}{\lambda^4}{\left(\frac{\delta T}{T}\right)}^2,
\end{eqnarray}
where $\ell_\gamma=1/\sigma_{\rm Th}n_e x_e$ is the mean free path of
the photons and $\sigma_{\rm Th}=\frac{8\pi}{3}\frac{\hbar^2
  \alpha^2}{m^2_e c^2}$ is the Thomson cross-section and $\lambda$ the wave-length of the perturbation. $\ell_\gamma$ is
directly linked to the interaction rate $\Gamma_{\rm Th}$ of the
Thomson scattering as $\Gamma_{\rm Th}=1/\ell_\gamma\propto
T^{3}$ and from CMB observations, we have the constraint $\frac{\delta
  T}{T}\approx 3\times10^{-5}$. 

Couple of points are worth noting regarding Eq.~\ref{omega}: first, by including the second-order terms, the estimate of the 
primordial vorticity has been improved by factor $4$ as compared to Ref. \cite{berezhiani2004generation}. Second, the vorticity ($\Omega$) is 
inversely proportional to the wavelength of the perturbation $\lambda$. 
Physically this is related to the fact that photons tend to diffuse
from the denser to rarer region and since the photons and electrons 
are tightly coupled the photons tend to carry along electrons leading to 
decrease in the strength of the magnetic field. In the next subsection, 
we fix the length scale to the Silk damping scale.

\subsection{Conductivity}\label{ss_conductivity}

One of the crucial steps in our analysis is that the process generating
vorticity do not necessarily contribute significantly to the generation of the 
seed magnetic fields. As we have shown in \ref{vorticities}, 
Thomson scattering contributes to the generation of vorticity, however, 
we show that the Coulomb scattering contributes to the conversion of 
vorticity to seed magnetic field.  

The conductivity $\kappa$ of the plasma is given by $\kappa=J/E$.  
The ratio of the two conductivities, $\kappa_{\rm Th}$ and $\kappa_{\rm Coul}$, is 
given by \cite{ahonen1996electrical, baym1997electrical}
\begin{eqnarray}\label{kappa}
\frac{\kappa_{\rm Th}}{\kappa_{\rm Coul}}&=&\frac{\beta\ln(\Lambda)}{\sqrt{2\pi}} 
\left(\frac{m_e c^2}{k_B T}\right)^{5/2}\\
&\approx&10^{-9}\times\left(T_{\rm MeV}\right)^{-5/2},\nonumber
\end{eqnarray}
where $\ln(\Lambda)\approx 10$ is the Coulomb logarithm,
$m_e=0.51$~MeV$/c^2$ the mass of the electron and $k_B$ the Boltzmann
constant (for details, see \ref{appendix_conductivity}). For
$T<100$~eV, $1/\kappa_{\rm Coul}>10/\kappa_{\rm Th}$. As a
consequence, in the temperature range considered, the contribution for the
conductivity $\kappa$ in the magneto-hydrodynamic limit in
Eq.~(\ref{MHD}) to generate magnetic fields is
\begin{eqnarray}
\kappa\approx\kappa_{\rm Coul}=6\epsilon^2_0\frac{\sqrt{2}
\left(\pi k_B T\right)^{3/2}}{m_e^{1/2} e^2 \ln(\Lambda)}.
\end{eqnarray}
For details, see also \ref{eq_MHD}.

\subsection{Fluctuations and coherence length}\label{ss_fluctuations}

Unlike the generation of the primordial density (scalar) perturbations, 
the generation of primordial magnetic field crucially rests on the coherence 
length. If the coherence length is small, then the net magnetic field may 
effectively decrease to zero and the generation mechanism is unsatisfactory. 

To obtain the relation, let us consider scalar perturbations of wave-length 
$\lambda_i$ entering the Hubble radius at time $t_i$ and temperature 
$T_i \gg 1$~eV. These perturbations produce vorticity in the plasma 
[cf. Eq.~(\ref{omega})] which 
eventually leads to the generation of the magnetic field. At time $t$ 
corresponding to a temperature $T$ (in the range $20~{\rm eV} < T < 
100~{\rm eV}$), seed magnetic field of strength $B_{\rm seed}$ and coherence length $\lambda$ is generated [cf. Eq.~(\ref{omega})]. As mentioned earlier, 
photons tend to diffuse from the denser to rarer region and since the 
photons and electrons are tightly coupled the photons tend to carry along 
electrons leading to the decrease in the strength of the magnetic field.
Hence, we set the wavelength at the time of generation of the seed magnetic 
field to be equal to the diffusion length of the photon. In other words,
we are looking at wave-lengths that are not affected by the Silk damping \cite{silk1968cosmic}. Under these assumptions, we get :
\begin{eqnarray}\label{L1}
\frac{ct\ell_\gamma}{\lambda}&=&\frac{3}{2},\\\label{L2}
\left(\frac{\ell_\gamma}{\lambda}\right)^2&=&
2\times10^{-3}\lambda^{2/3}_M,\\\label{L3}
\left(\frac{1~{\rm eV}}{T}\right)^{1/2}&=&0.85\times\lambda^{1/3}_M,
\end{eqnarray}
where $\lambda_M=\lambda_0/(1~{\rm Mpc})$ and $\lambda_0$ is the 
wavelength of the scalar perturbation today. Physically, $\lambda_M$ 
encodes the current coherence length of the magnetic field in Mpc. 
For detailed computation, see \ref{appendix_fluctuations}. 
 
\subsection{Seed magnetic field}\label{resulting_B}

Substituting Eqs.~(\ref{omega}) and (\ref{kappa}) in Eq.~(\ref{BMHD}),
we have
\begin{eqnarray}
\frac{\dd B_{\rm seed}}{{\left(k_B T\right)}^2}
\approx\frac{1.7\beta\ln(\Lambda)}{\sqrt{\epsilon_0 \hbar^3 c^5}}{
\left(\frac{m_e c^2}{k_B T}\right)}^{1/2}{\left(\frac{\delta T}{T}\right)}^2 
\frac{\ell_\gamma^3}{\lambda^4}\frac{\delta n_{ep}}{n}c\dd t\nonumber\\
\end{eqnarray}
Then, substituting the corresponding expressions with Eqs.~(\ref{L1}),
(\ref{L2}) and (\ref{L3}) and integrating from $T_i$ to $T$
(cf. \ref{appendix_fluctuations}), we can express the
strength of ${B}_{\rm seed}$ either as a function of the current
coherence scale $\lambda_M$ of the field or as a function of the
temperature $T$ at which the seed is estimated
\begin{eqnarray}\frac{{B}_{\rm seed}}{1~{\rm G}}&\approx&10^{-52}\times {\lambda_M}^{-5/3},\\
\frac{{B}_{\rm seed}}{1~{\rm G}}&\approx&10^{-52}\times \left(\frac{T}{1~{\rm eV}}\right)^{5/2}.
\end{eqnarray}
At a coherence length of 10 kpc, our model predicts the seed magnetic
field strength to be $10^{-49}$~G, which is tiny.

\section{Comparison with earlier results in the literature}\label{comparison}

Generation of magnetic field with Eq.~(\ref{gopal-sethi}) with the other source term on similar coherence length $10~{\rm kpc}$ has been studied in Ref. \cite{gopal2005generation}, where the evolution of the electron-proton-photon plasma without 
using the long wave-length (MHD) is considered. They obtain magnetic strength of $B<10^{-30}$~G 
from small coherence scale to largest possible scale $\lambda\in[10~{\rm kpc};100~{\rm Mpc}]$. We can see that in both cases, the seed magnetic fields generated are extremely small. It will be very difficult to explain the galactic magnetic field with these processes.

Over the last two decades, there have been other proposals in the literature 
to generate seed magnetic fields around recombination.
For instance, non-linear evolution of primordial fluctuations was 
investigated in \cite{matarrese2005large}. In this reference, the authors obtained the seed 
magnetic field strength of $B\approx10^{-23}(\lambda/{\rm Mpc})^2$~G 
with a coherence-length $\lambda>{\rm 1~Mpc}$. An other work the same topic (\cite{kobayashi2007cosmological}) estimated the amplitude of the magnetic field to be $B\approx10^{-27}$~G at recombination on horizon scale. In Ref.~\cite{fenu2011seed}, the authors 
obtained magnetic field strength of $B\approx10^{-29}(\lambda/{\rm Mpc})^2$~G
with a coherence-length $\lambda>{\rm 1~Mpc}$. Another process that has been studied 
is the generation of magnetic fields based on second order cosmological 
perturbation theory. In Ref. \cite{takahashi2005magnetic} for example, the 
authors estimated the strength of the magnetic field to be 
$B\approx10^{-19}(\lambda/{\rm Mpc})^2$~G for coherence length 
$\lambda\approx10~{\rm Mpc}$. As the reader will be able to notice, the 
above models generate magnetic field over a larger coherence length 
compared to our model. In \cite{ichiki2006cosmological}, it was claimed that the generation of seed magnetic fields of strength $B\approx 10^{-14}$ at 10 kpc coherence scale could be explained by second order perturbation theory but this allegation was later refuted in \cite{ichiki2007magnetic} and latter studies in \cite{saga2015magnetic} estimate seed magnetic field of amplitude of $10^{-24}$~G and coherence length of 1~Mpc.

However, even if the mechanism described here seems to be inefficient to produce seed magnetic field strong enough to explain the magnetic fields in large cosmological structures, \cite{banik2015recombination} follows an analogous approach with a different seed of vorticity arising from condensate of axions. They are able to generate seeds of magnitude $10^{-23}$~G on coherence length scale 10 kpc. As a consequence, we can expect that a modification of the seed vorticity could produce magnetic fields whose amplitude are big enough.

\section{Results and discussion}\label{conclusion}

\begin{figure}
\begin{tabular}{|c|l|c|}
\hline
References                       & Coherence length and strength                \\
%   & Models  \\
\hline\hline
%\cite{harrison1973magnetic}       & primordial             \\%&               \\
%\hline
%\cite{subramanian1998microwave}   & inhomogeneous $B$      \\%&               \\
%\hline
%Subramanian-Barrow 2001           & inhomogeneous $B$      \\%&               \\
%\hline
%Subramanian-Barrow 2002           & inhomogeneous $B$      &              \\
\hline
\cite{berezhiani2004generation}   & $\lambda<$ Mpc, $B\approx 10^{-23}$ G \\%& same as ours           \\
\hline
%Subramanian-Seshardi-Barrow 2003  & inhomogeneous $B$      &               \\
%\hline
%\cite{banerjee2004evolution}      & evol of stochastic $B$  \\%&               \\
\hline
\cite{matarrese2005large}         & $\lambda>1$ Mpc, $B\approx10^{-23}(\lambda/{\rm Mpc})^2$ G \\%& NL evolution of prim density perturbation grav. induce vorticity  \\
\hline
\cite{takahashi2005magnetic}      & $\lambda\approx10$ Mpc and $B\approx10^{-19}$~G \\%& generation of $B$ in 2nd order of cosmological perturbations \\
\hline
\cite{gopal2005generation}        & $100 {\rm Mpc}>\lambda>10 {\rm kpc}$, $B<10^{-30}$~G \\%& in pre-recomb era, generation of $B$ in 2nd order in perturbation theory  \\
%\hline
%\cite{ichiki2006cosmological}     & (1 Mpc,$10^{-18.1}$~G) and (10 kpc,$10^{-14.1}$~G) \\%& 2nd order coupling produce $B$ before recomb  \\
\hline
\cite{seshadri2009cosmic}         &  $B<35nG$\\%& primordial $B$ lead to NG signature in CMB              \\
\hline
\cite{fenu2011seed}               & $\lambda>1~Mpc$, $B\approx 3\times 10^{-29}$~G \\%& NL dynamics creates vortical currents when the tight-coupling approx breaks down around recombination. Generates $B$ at second order in cosmological perturbations.                                                             \\
\hline
\cite{paoletti2011cmb}             & $\lambda\approx1~Mpc$, $B<$~nG\\%& stochastic background of primordial $B$  \\
\hline
\cite{ade2013planck}               & $\lambda\approx1~Mpc$, $B<$~nG\\%& stochastic background of primordial $B$  \\
\hline
\cite{planckplanck}                & $\lambda\approx1~Mpc$, $B<$~nG\\%& stochastic background of primordial $B$  \\
\hline
 \cite{saga2015magnetic}           & $\lambda\approx 2~{\rm Mpc}$, $B_{\rm rec}=5\times 10^{-24}$~G \\%& B at recombination revisited  \\
\hline
  \cite{banik2015recombination}    & $\lambda\approx 10~{\rm kpc}$, $B_{\rm rec}=5\times 10^{-23}$~G\\%& B at recomb from axion DM \\
\hline
\label{different-models}

\end{tabular}
\caption{Summary of different references presenting different mechanisms to generate seed magnetic fields and their coherence length.}
\end{figure}

We discussed here a model to explain galactic and inter-galactic
large coherence-length magnetic fields. We considered the generation
of seed magnetic fields by the vorticity of the primordial
cosmological plasma just before recombination, at $T<100$~eV. We
demonstrated that the electrical conductivity in the MHD limit 
is dominated by the Coulomb scattering. 

We have emphasized that electrical resistivity dissipation time 
scales \cite{teodoro2008cautionary} are several orders of magnitude 
larger than the age of the Universe (\ref{diffusion}). 
This result is compatible with Ref.~\cite{fenu2011seed}. This implies 
that the magnetic fields generated in this process are sustainable and 
are not erased by resistive diffusion. Our analysis also shows that 
magneto-genesis within the frame-work of MHD is consistent. Since 
the MHD limit corresponds to large wave-lengths, the smaller 
wavelength (large frequency)  electromagnetic waves do not propagated. 
The dominating terms are those related to diffusion and amplification \cite{teodoro2008cautionary}. 

%%%
We have explicitly shown that the strength of the seed magnetic fields generated before recombination is $10^{-49}$~G on the current coherence scale of 10~kpc. We emphasize that our mechanism generates seed magnetic fields on small scales, whereas the main mechanisms studied in the literature are on scale bigger than 1 Mpc. This scale is directly due to the wavelength of primordial perturbation entering the horizon during radiation domination. The advantage of small scale magnetic fields is that they are easier to amplify with galactic dynamo, as we discuss it later in this Section. Compared to more exotic theories generating seed magnetic fields on similar scales, the strength of our fields are generally smaller. However, there are still different options to improve the strength of the seed magnetic fields especially with regards to the vorticity which is needed to generate them. 
%%%%%

It is difficult to compare
our model with current observations of galactic and extra-galactic magnetic fields on
kpc coherence scale. Based on FERMI observations of TeV sources at
$z\approx 0.1$, lower limit on inter-galactic magnetic fields $B_{\rm min}$ 
were established to be $B_{\rm min}=10^{-18}-10^{-15}$~G at
large scales in voids (see Refs. \cite{neronov2010evidence,tavecchio2010intergalactic}). The
observational constraints of large-scale magnetic fields in voids give an estimation
of the seed magnetic fields. It is important to point that some authors have critized the latter tests. 
In Ref. \cite{broderick2012cosmological} it was argued that the lack of an
inverse Compton GeV bump cannot be used as a constraint on the
inter-galactic magnetic fields because of plasma
instabilities. However, the presence of plasma instabilities in TeV
blazars beams involve an upper constraint on the inter-galactic
magnetic fields: $B\leq 10^{-12}$~G. Our results are
compatible with this constraint.

In \ref{amplification}, we have considered a more
realistic process for galactic dynamo amplification. After the first
stage of galaxy formation our model predicts $B\approx10^{-47}$~G. 
To explain the current observed magnetic fields in the galaxies 
and clusters of the order of $10^{-6}$~G coherent on scales of about 
$10~{\rm kpc}$, our model requires a huge galactic amplification. 
For massive galaxies, like the Milky Way, it is possible to have large 
amplification, however for small and/or lighter galaxies, like dwarf
galaxies \cite{dubois2009magnetised}, the current dynamo models 
do not provide large amplification. Also, the dynamo mechanism 
is not efficient in the inter-galactic medium. We have taken into account a dynamo effect as 
realistic as possible, but there is still room for improvement as 
the different dynamo effects still present numerous unknown features.

%%%%%%%%%%%
\section{Acknowledgments}

The authors would like to thank A.~Dolgov, Y.~Dubois, R.~Marteens, S.~Sethi and K.~Subramanian for useful discussions. The work is supported by Max Planck-India Partner Group on Gravity and Cosmology. SS is partially supported by
Ramanujan Fellowship of DST, India.


%% References


\bibliographystyle{elsarticle-num}
\bibliography{biblio_Bfields}






% \end{thebibliography}

\appendix

\section{Plasma equations}\label{plasma_equations}

From \cite{takahashi2008electromagnetic}, we have the following equations

\begin{eqnarray}\label{eq_electron}
\partial_t \vec{v}_e+ (\vec{v}_e.\vec{\nabla}) \vec{v}_e +H\vec{v}_e &=& -\frac{\vec{\nabla} P_e}{\rho_e} -\frac{e}{m_e}(\vec{E}+\vec{v}_e\times\vec{B})\nonumber\\
   &+&\frac{1}{m_e n_e}\left[\frac{e^2n_e n_p}{\kappa_{\rm Coul}}(\vec{v}_p-\vec{v}_e)+\frac{\rho_\gamma}{\tau_{e\gamma}}(\vec{v}_\gamma-\vec{v}_e-\frac{1}{4}\vec{v}_e.\Pi_\gamma)\right]-\vec{\nabla}\phi\nonumber\\ \\\label{eq_proton}
\partial_t \vec{v}_p+ (\vec{v}_p.\vec{\nabla}) \vec{v}_p +H\vec{v}_p &=& -\frac{\vec{\nabla} P_p}{\rho_p}+\frac{e}{m_p}(\vec{E}+\vec{v}_p\times\vec{B})\nonumber\\
  &+&\frac{1}{m_p n_p}\left[\frac{e^2n_e n_p}{\kappa_{\rm Coul}}(\vec{v}_e-\vec{v}_p)+\left(\frac{m_e}{m_p}\right)^2\frac{\rho_\gamma}{\tau_{e\gamma}}(\vec{v}_\gamma-\vec{v}_p-\frac{1}{4}\vec{v}_p.\Pi_\gamma)\right]-\vec{\nabla}\phi.\nonumber\\
\end{eqnarray}
where $\{\vec{v}_i, P_i, \rho_i, n_i\}$ are the velocity, pressure, energy density and particle density of the fluid $i$, $\Pi_{\gamma}$ the tensor of anisotropic pressure of the photons, $\phi$ the gravitational potential, $\kappa_{\rm Coul}$ the conductivity of the medium (due to Coulomb scattering), $\tau_{\gamma e}$ the typical time between two shocks of a photon on electrons (Thomson scattering), and $\vec{E}$ and $\vec{B}$ the electric and magnetic fields in the plasma. Subtracting Eq.~(\ref{eq_electron}) to Eq.~(\ref{eq_proton}), we can deduce
\begin{eqnarray}
\partial_t \delta \vec{v}_{pe}&+&\left[\left(\vec{v}.\vec{\nabla}\right)\delta \vec{v}_{pe}+\left(\delta \vec{v}_{pe}.\vec{\nabla}\right)\delta \vec{v}\right]+H\delta \vec{v}_{pe}\approx e\left(\frac{1}{m_p}+\frac{1}{m_e}\right)\vec{E}\nonumber\\
 &+&e\left(\frac{\vec{v}_{p}}{m_p}+\frac{\vec{v}_{e}}{m_e}\right)\times\vec{B}-\frac{e^2n}{\kappa_{\rm Coul}}\left(\frac{1}{m_p}+\frac{1}{m_e}\right)\delta\vec{v}_{pe}\nonumber\\
                              &+&\left(\frac{m_e}{m_p}\right)^2\frac{1}{m_p n_p}\frac{\rho_\gamma}{\tau_{e\gamma}}\left[\vec{v}_\gamma-\vec{v}_p-\frac{1}{4}\vec{v}_p.\Pi_\gamma\right]+\frac{1}{m_e n_e}\frac{\rho_\gamma}{\tau_{e\gamma}}\left[\vec{v}_e-\vec{v}_\gamma+\frac{1}{4}\vec{v}_e.\Pi_\gamma\right]-\frac{\vec{\nabla} P_p}{\rho_p}+\frac{\vec{\nabla} P_e}{\rho_e}.\nonumber\\
\end{eqnarray}
The term in $\left(\frac{m_e}{m_p}\right)^2$ ($\approx 10^{-6}$) for protons can be considered as negligible in front of the analog term for electrons. We also neglect the quantity $1/m_p$ compared to $1/m_e$, and $\vec{v}_p/m_p$ compared to $\vec{v}_e/m_e$. In addition to the previous simplifications, for $i\in\{e,p\}$
\begin{eqnarray}
\frac{\vec{\nabla}P_i}{\rho_i}=\frac{1}{m_i}\left(\vec{\nabla}T+T\frac{\vec{\nabla}n_i}{n_i}\right).
\end{eqnarray}
At $0^{\rm th}$ order, $n_e\approx n_p$ so that $\frac{\vec{\nabla}n_e}{n_e}\approx \frac{\vec{\nabla}n_p}{n_p}$. With $m_e/m_p\approx10^{-3}$, then
we have $||\frac{\vec{\nabla} P_p}{\rho_p}||\ll ||\frac{\vec{\nabla} P_e}{\rho_e}||$. As a consequence,
\begin{eqnarray}
\partial_t \delta \vec{v}_{pe}+\left[\left(\vec{v}.\vec{\nabla}\right)\delta \vec{v}_{pe}+\left(\delta \vec{v}_{pe}.\vec{\nabla}\right)\delta \vec{v}\right]+H\delta \vec{v}_{pe}\approx \frac{\vec{\nabla} P_e}{\rho_e}&+&\frac{e}{m_e}\left(\vec{E}+\vec{v}_{e}\times\vec{B}\right)-\frac{e^2n}{m_e\kappa_{\rm Coul}}\delta\vec{v}_{pe}\nonumber\\&+&\frac{1}{m_e n_e}\frac{\rho_\gamma}{\tau_{e\gamma}}\left[\vec{v}_e-\vec{v}_\gamma+\frac{1}{4}\vec{v}_e.\Pi_\gamma\right].
\end{eqnarray}
Knowing that $\frac{\vec{J}}{en_e}\approx\delta\vec{v}_{pe}$,
\begin{eqnarray}
\partial_t \left[\frac{\vec{J}}{en_e}\right]+\left[\left(\vec{v}.\vec{\nabla}\right)\left[\frac{\vec{J}}{en_e}\right]+\left(\left[\frac{\vec{J}}{en_e}\right].\vec{\nabla}\right)\delta \vec{v}\right]+H\left[\frac{\vec{J}}{en_e}\right]\approx\frac{\vec{\nabla} P_e}{\rho_e}&+&\frac{e}{m_e}\left(\vec{E}+\vec{v}_{e}\times\vec{B}\right)-\frac{e^2n}{m_e\kappa_{\rm Coul}}\left[\frac{\vec{J}}{en_e}\right]\nonumber\\
&+&\frac{1}{m_e n_e}\frac{\rho_\gamma}{\tau_{e\gamma}}\left[\vec{v}_e-\vec{v}_\gamma+\frac{1}{4}\vec{v}_e.\Pi_\gamma\right].
\end{eqnarray}
Let us take the curl of the previous equation. Knowing the equation of Maxwell-Faraday $\vec{\nabla}\times\vec{E}=-\partial_t \vec{B}$, we thus have
\begin{eqnarray}
\vec{\nabla}\times\partial_t \left[\frac{\vec{J}}{en_e}\right]&+&\vec{\nabla}\times\left[\left(\vec{v}.\vec{\nabla}\right)\left[\frac{\vec{J}}{en_e}\right]+\left(\left[\frac{\vec{J}}{en_e}\right].\vec{\nabla}\right)\delta \vec{v}\right]+\vec{\nabla}\times H\left[\frac{\vec{J}}{en_e}\right]\approx \vec{\nabla}\times\left(\frac{\vec{\nabla} P_e}{\rho_e}\right)
-\frac{e}{m_e}\partial_t \vec{B}\nonumber\\&+&\frac{e}{m_e}\vec{\nabla}\times\left(\vec{v}_{e}\times\vec{B}\right)-\vec{\nabla}\times\frac{e^2n}{m_e\kappa_{\rm Coul}}\left[\frac{\vec{J}}{en_e}\right]+\vec{\nabla}\times\frac{1}{m_e n_e}\frac{\rho_\gamma}{\tau_{e\gamma}}\left[\vec{v}_e-\vec{v}_\gamma+\frac{1}{4}\vec{v}_e.\Pi_\gamma\right]
\end{eqnarray}
The terms on the left side of the equation are very small compared to $||\frac{e}{m_e}\partial_t \vec{B}||$ because 
\begin{eqnarray}
\frac{\frac{e}{m_e}\partial_t \vec{B}}{v \frac{\vec{J}}{en_e}}&\approx&\frac{\frac{e}{m_e}\partial_t \vec{B}}{k \partial_t \left[\frac{\vec{J}}{en_e}\right]}\\
&\approx& \frac{t}{\tau_{ep}}\gg1
\end{eqnarray}
We can finally write the following equation ruling the behavior of magnetic fields in the plasma
\begin{eqnarray}
\partial_t \vec{B}\approx\vec{\nabla}\times\left(\vec{v}_e \times \vec{B}\right)+\frac{1}{e}\vec{\nabla}\times\left(\frac{\vec{\nabla} P_e}{n_e}\right)-\vec{\nabla}\times\frac{m_e}{\tau_{ep}e^2}\left[\frac{\vec{J}}{n_e}\right]-\vec{\nabla}\times\frac{Rm_e}{e\tau_{\gamma e}}\left(\delta\vec{v}_{\gamma e}-\frac{1}{4}\vec{v}_e.\Pi_\gamma\right),\nonumber\\
\end{eqnarray}
where $R=\rho_\gamma/n_e m_e$.

\section{Computation of vorticity: upto second order in distribution function }\label{appendix_vorticity}

In this Appendix, we provide detailed computation of the derivation of 
the vorticity in Eq.~(\ref{omega}) by expanding the Boltzmann equation 
for the photon distribution $f_\gamma$ up to the second order. The Boltzmann equation for the photon distribution $f_\gamma$ is given by
\begin{eqnarray}
  \left(\partial_t +\vec{V}.\nabla -H\vec{p}.\partial_{\vec{p}} 
+\vec{F}.\partial_{\vec{p}} \right)f_\gamma (t,\vec{x},E,\vec{p})=I_{\rm coll} [f_i],
\end{eqnarray}
where $V_k$ is the photon velocity. It is important to note that the 
product of an odd number of $V_k$ is equal to zero whereas 
$V_k V_i=\frac{1}{3}$. 

Let us define the operator
$\mathcal{K}=\partial_t +\vec{V}.\nabla -H\vec{p}.\partial_{\vec{p}}
+\vec{F}.\partial_{\vec{p}}$. In our analysis we have ignored the effects 
of gravity (cosmological expansion) as we are working on time scales such 
that the expansion does not drastically 
affects the results. We also assume that there is no other external force 
acting on the plasma. As a consequence, $\mathcal{K}=\partial_t
+\vec{V}.\nabla$. The velocity of the fluid is given by

\begin{eqnarray}
v_k=\frac{1}{\int\dd^3 \vec{p}~f_\gamma^{(0)}}\int\dd^3 \vec{p}~V_k f_\gamma~.
\end{eqnarray}

For the temperature range considered, the Thomson scattering rate dominates 
the collision term in the RHS of the Boltzmann equation. 
\begin{eqnarray}
\Gamma_{\rm Th}&\propto& T^3\propto \ell^{-1}_{\gamma}\\
\partial_t \Gamma_{\rm Th}&=& 3\Gamma_{\rm Th} \frac{\partial_t T}{T}\\
\partial_i \Gamma_{\rm Th}&=& 3\Gamma_{\rm Th} \frac{\partial_i T}{T}
\end{eqnarray}
We assume that the elastic electron-photon rate is high and that the
integrals over time are dominated by small values of time $\tau$. We
can thus expand $\Gamma_{\rm Th}(t-\tau_2,\vec{x}-\vec{v}\tau_2 )$ as
following

\begin{eqnarray}
  \Gamma_{\rm Th}(t-\tau_2,\vec{x}-\vec{v}\tau_2 )&\approx& 
\Gamma_{\rm Th}(t,\vec{x})-\tau_2 \left[\partial_t 
\Gamma_{\rm Th}(t,\vec{x})+V_i \partial_i \Gamma_{\rm Th}(t,\vec{x})\right]
\nonumber\\
\exp\left(-\int_0^{\tau_1}\dd \tau_2 \Gamma_{\rm Th}(t-\tau_2,\vec{x} - 
\vec{v}\tau_2 )\right)&\approx&e^{-\tau_1  \Gamma_{\rm Th}(t,\vec{x})}
\times\left(1+\frac{3}{2}\tau^2_1 \Gamma_{\rm Th}(t,\vec{x})
\left[\frac{\partial_t T}{T}+V_i \frac{\partial_i T}{T}\right]\right)\nonumber\\
\end{eqnarray}

\subsection{$0^{\rm th}$ order: $f_\gamma\approx f_\gamma^{(0)}$}

At $0^{\rm th}$ order, $f_\gamma^{(0)}=e^{-E/T}$ and there is no
collision term $I_{\rm coll}[f_i]$ in the Boltzmann equation

\begin{eqnarray}
\mathcal{K}f_\gamma^{(0)}=0
\end{eqnarray}

\begin{eqnarray}
v^{(0)}_k&=&\frac{1}{\int\dd^3 \vec{p}~f_\gamma^{(0)}}\int\dd^3 
\vec{p}~V_k f_\gamma^{(0)}\\
         &=& \frac{1}{\int\dd^3 \vec{p}~f_\gamma^{(0)}}\int\dd^3 
\vec{p}~V_k e^{-E/T}=0
\end{eqnarray}

$$\boxed{v^{(0)}_k=0}$$

As a consequence, the $0^{\rm th}$ order, no vorticity is generated in
the fluid.

\subsection{$1^{\rm st}$ order: $f_\gamma\approx f_\gamma^{(0)}+f_\gamma^{(1)}$}

At first order, the collision term is no longer equal zero in the Boltzmann equation

\begin{eqnarray}
\mathcal{K}\left[f_\gamma^{(0)}+f_\gamma^{(1)}\right] &=&-\Gamma_{\rm Th} 
f_\gamma^{(1)}\\\label{f1}
\left(\mathcal{K}+\Gamma_{\rm Th}\right)f_\gamma^{(1)}&=&-\mathcal{K}f_\gamma^{(0)}.
\end{eqnarray}

Solving Eq.~(\ref{f1}), we obtain

\begin{eqnarray}
f_\gamma^{(1)}&=&-\int_0^{t}\dd \tau_1 \exp\left[-\int_0^{\tau_1}\dd \tau_2 
\Gamma_{\rm Th} (t-\tau_2,\vec{x}-\vec{v}\tau_2)\right]\times\mathcal{K}
f_\gamma^{(0)} (t-\tau_1,\vec{x}-\vec{v}\tau_1)
\end{eqnarray}

We expand $f_\gamma^{(0)}(t-\tau_1,\vec{x}-\vec{v}\tau_1)$ in the same
way as we expanded $\Gamma(t-\tau_1,\vec{x}-\vec{v}\tau_1)$

\begin{eqnarray}
f_\gamma^{(0)} (t-\tau_1,\vec{x}-\vec{v}\tau_1)\approx 
f_\gamma^{(0)} (t,\vec{x})-\tau_1 \left(\partial_t f_\gamma^{(0)} (t,\vec{x})
+V_i \partial_i f_\gamma^{(0)} (t,\vec{x})\right).
\end{eqnarray}

Let us now expand the exponential function in $\tau\Gamma$ and
integrate over time

\begin{eqnarray}
f_\gamma^{(1)}&=&-\int_0^{t}\dd \tau_1 e^{-\tau_1  
\Gamma_{\rm Th}(t,\vec{x})}\times\left(1+\frac{3}{2}\tau^2_1 
\Gamma_{\rm Th}(t,\vec{x})\left[\frac{\partial_t T}{T} + 
V_i \frac{\partial_i T}{T}\right]\right)\times\left[\mathcal{K}f_\gamma^{(0)} 
- \tau_1(\partial_t+V_j\partial_j)\mathcal{K}f_\gamma^{(0)}\right]\nonumber\\
&=&-\Big(\frac{1}{\Gamma_{\rm Th}}\mathcal{K}f_\gamma^{(0)} - 
\frac{1}{\Gamma_{\rm Th}^2}(\partial_t+V_j\partial_j)\mathcal{K}f_\gamma^{(0)}
+\frac{3}{\Gamma_{\rm Th}^2}\left[\frac{\partial_t T}{T}+V_i 
\frac{\partial_i T}{T}\right]\mathcal{K}f_\gamma^{(0)}\Big)
\end{eqnarray}

\begin{eqnarray}
f_\gamma^{(1)}=&-&e^{-E/T}V_i\Big(\frac{1}{\Gamma_{\rm Th}}
\frac{E}{T}\frac{\partial_i T}{T}-\frac{2}{\Gamma_{\rm Th}^2} 
\frac{E}{T}\frac{\partial_t \partial_i T}{T}
-\frac{2}{\Gamma_{\rm Th}^2} \frac{E^2}{T^2}\frac{\partial_t T}{T}\frac{\partial_i T}{T}
+\frac{10}{\Gamma_{\rm Th}^2}\frac{E}{T}\frac{\partial_t T}{T}\frac{\partial_i T}{T}\Big)\nonumber\\
&-&e^{-E/T}\Big(\frac{1}{\Gamma_{\rm Th}}\frac{E}{T}\frac{\partial_t T}{T}
-\frac{1}{\Gamma_{\rm Th}^2}\frac{E^2}{T^2}{\left(\frac{\partial_t T}{T}\right)}^2
-\frac{1}{\Gamma_{\rm Th}^2}\frac{E}{T}\frac{\partial_{tt} T}{T}
+\frac{5}{\Gamma_{\rm Th}^2}\frac{E}{T}{\left(\frac{\partial_t T}{T}\right)}^2
\nonumber\\
&-&\frac{1}{\Gamma_{\rm Th}^2}V_j V_i \frac{E^2}{T^2}
\frac{\partial_i T}{T}\frac{\partial_j T}{T}-\frac{1}{\Gamma_{\rm Th}^2}V_j 
V_i\frac{E}{T}\frac{\partial_{ij} T}{T}
+\frac{5}{\Gamma_{\rm Th}^2}V_j V_i\frac{E}{T}\frac{\partial_i T}{T}
\frac{\partial_j T}{T} \Big)
\end{eqnarray}

Finally, we can compute the fluid velocity at first order $v^{(1)}$

\begin{eqnarray}
v^{(1)}&=&\frac{1}{\int\dd^{3}\vec{p} f_\gamma^{(0)}}\int\dd^{3}\vec{p}V_k 
f_\gamma^{(1)}\\
&=&-\Big(\frac{1}{\Gamma_{\rm Th}}\frac{\partial_k T}{T}
-\frac{2}{\Gamma_{\rm Th}^2}\frac{\partial_t \partial_k T}{T}
-\frac{8}{\Gamma_{\rm Th}^2}\frac{\partial_t T}{T}\frac{\partial_k T}{T}
+\frac{10}{\Gamma_{\rm Th}^2}\frac{\partial_t T}{T}\frac{\partial_k T}{T}\Big)
\end{eqnarray}

$$\boxed{v^{(1)}=-\frac{1}{\Gamma_{\rm Th}}\frac{\partial_k T}{T}
+\frac{2}{\Gamma_{\rm Th}^2}\left(\frac{\partial_t \partial_k T}{T}
-\frac{\partial_t T}{T} \frac{\partial_k T}{T}\right)}$$

\subsection{$2^{\rm nd}$ order: $f_\gamma\approx f_\gamma^{(0)}+f_\gamma^{(1)}+f_\gamma^{(2)}$}

Proceeding in the same way as above, the second order expansion of the Boltzmann equation is given by:
\begin{eqnarray}
(\mathcal{K}+\Gamma_{\rm Th})f_\gamma^{(2)}&=&-\mathcal{K}f_\gamma^{(1)}\nonumber\\
f_\gamma^{(2)}&=&-\int_0^{t}\dd \tau_1 \exp\left[-\int_0^{\tau_1}\dd \tau_2 
\Gamma_{\rm Th} (t-\tau_2,\vec{x}-\vec{v}\tau_2)\right]\mathcal{K}
f_\gamma^{(1)}(t-\tau_1,\vec{x}-\vec{v}\tau_1 )\nonumber
\end{eqnarray}

Expand $f_\gamma^{(1)} (t-\tau_1,\vec{x}-\vec{v}\tau_1)$ as
we did earlier for $f_\gamma^{(0)}(t-\tau_1,\vec{x}-\vec{v}\tau_1)$
and $\Gamma(t-\tau_1,\vec{x}-\vec{v}\tau_1)$

\begin{eqnarray}
  f_\gamma^{(1)} (t-\tau_1,\vec{x}-\vec{v}\tau_1)\approx f_\gamma^{(1)} (t,\vec{x})
-\tau_1 \left(\partial_t f_\gamma^{(1)} (t,\vec{x})+V_i \partial_i 
f_\gamma^{(1)} (t,\vec{x})\right).
\end{eqnarray}

Neglecting contributions to the veloctity arising from the terms $\propto \frac{1}{\Gamma^{n}}$ with $n>2$, we get. 

\begin{eqnarray}
\partial_\alpha f^{(1)}=&-&\frac{1}{\Gamma_{\rm Th}}e^{-E/T}\Big(\frac{E^2}{T^2}
\frac{\partial_\alpha T}{T}\frac{\partial_t T}{T}
-5\frac{E}{T}\frac{\partial_\alpha T}{T}\frac{\partial_t T}{T}
+\frac{E}{T}\frac{\partial_\alpha \partial_t T}{T}\Big)\nonumber\\
                        &-&\frac{1}{\Gamma_{\rm Th}}e^{-E/T}V_i
\Big(\frac{E^2}{T^2}\frac{\partial_\alpha T}{T}\frac{\partial_i T}{T}
-5\frac{E}{T}\frac{\partial_\alpha T}{T}\frac{\partial_i T}{T}
+\frac{E}{T}\frac{\partial_\alpha \partial_i T}{T}\Big)
\end{eqnarray}

We can deduce from the former equation the contribution to fluid
velocity at second order in $1/\Gamma$

\begin{eqnarray}
v^{(2)}_k&=&\frac{1}{\int\dd^3 \vec{p} f^{(0)}_\gamma} \int \dd^3 \vec{p} 
V_k f^{(2)}_\gamma\\
&=&-\frac{1}{\int\dd^3 \vec{p} f^{(0)}_\gamma} \int \dd^3 \vec{p} V_k 
\int e^{-\tau_1 \Gamma_{\rm Th}}\mathcal{K}f^{(1)}_\gamma\\
&=&-\frac{1}{\int\dd^3 \vec{p} f^{(0)}_\gamma} \int \dd^3 \vec{p} V_k 
\int e^{-\tau_1 \Gamma_{\rm Th}}\left(\partial_t f^{(1)}_{\gamma} 
+ V_j \partial_j f^{(1)}_{\gamma}\right)\\
&=&\frac{1}{\Gamma_{\rm Th} T^3} \int \dd^3 \vec{p} V_k 
\int e^{-\tau_1 \Gamma_{\rm Th}}V_i\Big(\frac{E^2}{T^2}
\frac{\partial_t T}{T}\frac{\partial_i T}{T}
-5\frac{E}{T}\frac{\partial_t T}{T}\frac{\partial_i T}{T}
+\frac{E}{T}\frac{\partial_t \partial_i T}{T}\Big)\nonumber\\
&=&\frac{1}{\Gamma_{\rm Th}^2 3T^3} \Big(\frac{4! T^5}{T^2}
\frac{\partial_t T}{T}\frac{\partial_k T}{T}-5\frac{3! T^4}{T}
\frac{\partial_t T}{T}\frac{\partial_k T}{T}+\frac{3! T^4}{T}
\frac{\partial_t \partial_k T}{T}\Big)
\end{eqnarray}

Finally we obtain the following expression for $v^{(2)}$

$$\boxed{v^{(2)}_k=\frac{2}{\Gamma_{\rm Th}^2}  
\Big(\frac{\partial_t \partial_k T}{T}-\frac{\partial_t T}{T}
\frac{\partial_k T}{T}\Big)}$$

\subsection{Final expression}

\begin{eqnarray}
v_k&=&v_k^{(0)}+v_k^{(1)}+v_k^{(2)}\\
   &=&-\frac{1}{\Gamma_{\rm Th}}\frac{\partial_k T}{T}
+\frac{1}{\Gamma_{\rm Th}^2}\left[4\frac{\partial_k \partial_t T}{T}
-4\frac{\partial_k T \partial_t T}{T^2}\right]\\
\partial_j v_k&=&-\frac{24}{\Gamma_{\rm Th}^2}\frac{\partial_j T}{T}
\frac{\partial_k \partial_t T}{T}+~{\rm symmetric~terms~in~}j\leftrightarrow k
\end{eqnarray}

Due to the symmetry properties, we can notice that the term 
$\partial_i T/T$ does not contribute to vorticity. Using the fact that 
$\ell_{\gamma}\propto 1/n_e\approx T^{-3}$, we have

\begin{eqnarray}
\Omega_i &=& \epsilon_{ijk}\partial_j v_k\\
         &=& -24\times~ \ell^2_{\gamma} \epsilon_{ijk}\frac{\partial_j T}{T} \times \frac{\partial_k\partial_t T}{T}\\
\end{eqnarray}

We have the following equation of diffusion in the plasma 
(see \cite{berezhiani2004generation})

\begin{eqnarray}
\partial_t T = \frac{\ell_{\gamma} c}{3} \Delta T. 
\end{eqnarray}

Switching in Fourier space, we get

\begin{eqnarray}
\partial_t T = \frac{\ell_{\gamma} c}{3} k^2 T
\end{eqnarray}

As a consequence, we can estimate the vorticity

\begin{eqnarray}
\Omega_i  &\approx& 24c\times ~ 
\ell^2_{\gamma} \epsilon_{ijk}\frac{\partial_j T}{T} \times 
\frac{\partial_k (\frac{\ell_{\gamma}}{3} \Delta T)}{T}\nonumber\\
            &\approx& 24c\times~ \ell^2_{\gamma} 
\frac{\partial_j T}{T} \times \frac{ ( \frac{\ell_{\gamma}}{3} k^2 
\partial_k T)}{T}\nonumber\\
                &\approx& 8c\times~ \ell^3_{\gamma} k^4 
\left(\frac{\delta T}{T}\right )^2\nonumber
\end{eqnarray}

$$\boxed{\Omega_i \approx 12c\times 10^3 ~ 
\frac{\ell^3_{\gamma}}{ \lambda^4} \left (\frac{\delta T}{T}\right )^2}$$

By taking into account second order term in fluid velocity, the 
vorticity we have obtained is 4 times larger than the one that 
was previously obtained in Ref. \cite{berezhiani2004generation}. 
It is important to note that when the temperature decreases the 
fraction of free electrons also decrease,  and just before 
recombination the plasma is no longer at equilibrium and 
Eq.~(\ref{omega}) is no longer valid.

\section{Ratio of Vorticity generated due to Thomson and Coulomb scattering}\label{vorticities}

The vorticity generated in the Plasma by taking in to account the Thomson scattering  
is given by:
\begin{eqnarray}
\Omega_{\rm Th}&=&[...]\times\int_0^t \dd\tau~\tau\times \exp\left[-\int_0^\tau \dd\tau_2\Gamma_{\rm Th}\left(t-\tau_2,\vec{x}-\vec{V}\tau_2 \right)\right]\\
&\approx&[...]\times\ell_{\gamma}^2,
\end{eqnarray}
whereas the vorticity generated in the Plasma by taking in to account the  Coulomb scattering is given by
\begin{eqnarray}
\Omega_{\rm Coul}&=&[...]\times\int_0^t \dd\tau~\tau\times 
\exp\left[-\int_0^\tau \dd\tau_2\Gamma_{\rm Coul}\left(t-\tau_2,\vec{x} 
-\vec{V}\tau_2 \right)\right]\\
&\approx&[...]\times\ell_{e}^2.
\end{eqnarray}
We can now compute the ratio of the two vorticities

\begin{eqnarray}
\frac{\Omega_{\rm Th}}{\Omega_{\rm Coul}}&=&\left(\frac{\ell_{\gamma}}{\ell_{e}}\right)^2\\
&=&\frac{1}{3\beta^2}\frac{k_B T}{m_e c^2}\\
&\approx& 10^{20}\times T_{\rm MeV}
\end{eqnarray}

As a consequence, for the temperature range considered, $\Omega_{\rm
  Th} \gg \Omega_{\rm Coul}$, and Thomson scattering is the dominating
counterpart for the generation of vorticity in the plasma.

\section{Ratio of Thomson and Coulomb conductivities}\label{appendix_conductivity}

The equation of motion of an electron in an electric field is the
following

\begin{eqnarray}
m_e \frac{\dd \vec{V_e}}{\dd t}=-e\vec{E}.
\end{eqnarray}

The conductivity $\kappa=J/{E}$, where $J=-en_eV_e$, of the plasma is given by
\begin{eqnarray}
\kappa&=\frac{e^2 n_e \Delta t}{m_e}.
\end{eqnarray}
where $\Delta t$ is the typical time between two collisions. The
typical time between two electron-photon shocks is
(cf. \cite{berezhiani2004generation})
\begin{eqnarray}
{\Delta t}_{\rm Th}=\frac{\ell_e}{V_T},
\end{eqnarray}
where 
$$
\ell_e=c\sqrt{\frac{3m_e}{k_B T}}\frac{1}{\sigma_{\rm Th}n_\gamma}
$$ 
is the mean free path of the electron and $V_T=\sqrt{3k_B T/{m_e}}$ is the 
electron thermal velocity. The typical time between two electron-proton shocks is \cite{callen2003fundamentals}
\begin{eqnarray}
{\Delta t}_{\rm Coul}&=&\tau_{ep}\\
                     &=&\frac{6\sqrt{2} 
\epsilon^2_0 m_e^{1/2}\left(\pi k_B T\right)^{3/2}}{n_p e^4 \ln(\Lambda)}.
\end{eqnarray}

As a consequence, we can deduce the conductivity due to Thomson
scattering \cite{berezhiani2004generation}

\begin{eqnarray}
\kappa_{\rm Th}&=& \tau_{\gamma e}\\
               &=&\frac{3}{2}\frac{n_e m_e^2 }{n_{\gamma} k_B T }
\frac{c^4 \epsilon_0}{\alpha\hbar },
\end{eqnarray}

and the conductivity due to Coulomb scattering \cite{dinklage2005plasma}

\begin{eqnarray}
\kappa_{\rm Coul}=6\sqrt{2}\epsilon^2_0
\frac{n_e\left(\pi k_B T\right)^{3/2}}{m_e^{1/2} e^2 n_p \ln(\Lambda)}.
\end{eqnarray}

To estimate the most important contribution into the MHD equation,
let us evaluate the ratio of the two conductivities

\begin{eqnarray}
\frac{\kappa_{\rm Th}}{\kappa_{\rm Coul}}&=&\frac{\beta\ln(\Lambda)}{\sqrt{2\pi}} \left(\frac{m_e c^2}{k_B T}\right)^{5/2}\\
&\approx&10^{-9}\times \frac{1}{\left(T_{\rm MeV}\right)^{5/2}}
\end{eqnarray}

\begin{eqnarray}\label{bestK}
\Longrightarrow \kappa_{\rm Th}=\kappa_{\rm Coul} \Longleftrightarrow T=
250\mathrm{~eV~}
\end{eqnarray}

At $T_1=650$~eV, $1/\kappa_{\rm Th}=10/\kappa_{\rm Coul}$ and at
$T_2=100$~eV, $1/\kappa_{\rm Coul}=10/\kappa_{\rm Th}$. 

\section{Conductivity due to Coulomb scattering in the MHD equation}\label{eq_MHD}

\subsection{Coulomb vs Thomson}

In this Appendix, we show that the Coulomb scattering contributes 
significantly in the generation of magnetic field in the temperature range 
of our interest. Let us consider the following MHD equations for the two 
scattering processes:
\begin{eqnarray}
\partial_t \vec{B}_{\rm Th}&=&\nabla\times(\vec{v}\times\vec{B}_{\rm Th})
+\frac{1}{\kappa_{\rm Th}}\nabla\times\vec{J}\\
\partial_t \vec{B}_{\rm Coul}&=&\nabla\times(\vec{v}\times\vec{B}_{\rm Coul})
+\frac{1}{\kappa_{\rm Coul}}\nabla\times\vec{J}
\end{eqnarray}
The total contribution from the two processes can be approximately written as
\begin{eqnarray}
  \Longrightarrow\partial_t \left(\vec{B}_{\rm Th} + 
\vec{B}_{\rm Coul}\right)=\nabla\times\left[\vec{v}\times
\left(\vec{B}_{\rm Th}+\vec{B}_{\rm Coul}\right)\right]
+\left(\frac{1}{\kappa_{\rm Th}}
+\frac{1}{\kappa_{\rm Coul}}\right)\nabla\times\vec{J}\nonumber
\end{eqnarray}
we can define an effective magnetic field
\begin{eqnarray}
\vec{B}_{\rm seed}=\vec{B}_{\rm Th}+ \vec{B}_{\rm Coul}
\end{eqnarray}
and an effective conductivity
\begin{eqnarray}
\frac{1}{\kappa}&=&\frac{1}{\kappa_{\rm Th}}+ \frac{1}{\kappa_{\rm Coul}}
\end{eqnarray}
At $T_1=650$~eV, $1/\kappa_{\rm Th}=10/\kappa_{\rm Coul}$ and at
$T_2=100$~eV, $1/\kappa_{\rm Coul}=10/\kappa_{\rm Th}$. As a
consequence, the contribution for the conductivity $\kappa$ in the
magneto-hydrodynamic limit in Eq.~(\ref{MHD}) is

\begin{eqnarray}
\kappa\approx\left\{ \begin{array}{lll}
   \kappa_{\rm Coul}=6\epsilon^2_0\frac{\sqrt{2}
\left(\pi k_B T\right)^{3/2}}{m_e^{1/2} e^2 \ln(\Lambda)} \mathrm{~for~} 
T<100~{\rm eV}\\
   \kappa_{\rm Th}=\frac{3}{2}\beta\frac{m_e^2 }{k_B T }
\frac{c^4 \epsilon_0}{\alpha\hbar} \mathrm{~for~} T>650~{\rm eV}\\
   \kappa_{\rm Coul}\kappa_{\rm Th}/\left(\kappa_{\rm Coul}
+\kappa_{\rm Th}\right) \mathrm{~for~}T\in [100~{\rm eV};650~{\rm eV}] \\
   \end{array}\right.
\end{eqnarray}

As a consequence, the equation for the generation of the seed magnetic
field is

$$\boxed{\partial_t \vec{B}_{\rm seed}=\nabla\times(\vec{v}\times
\vec{B}_{\rm seed})+\frac{1}{\kappa}\nabla\times\vec{J}}$$

To be more rigorous, an additional term $HB_{\rm seed}$ should
be added on the right side of the equation. However, as the typical
time scale of the generation of the magnetic field is smaller than the
expansion time, it will not have much impact on the estimation of the
field and we can consider it as constant multiplying $B_{\rm seed}$
\cite{teodoro2008cautionary}.

\section{Expression of the seed magnetic field}

\begin{eqnarray}
B_{\rm seed}\approx\exp(-kvt)\left[\int_{t_i}^t \dd 
\tau \frac{2\pi J}{\lambda \kappa}\exp(kv\tau)+{\rm constant}\right]
\end{eqnarray}
where $kvt\approx 0.015~{\rm T}/1~{\rm eV}$ is a small quantity, so we can
consider that $\exp(-kvt)\approx 1$ and $\exp(-kvt)\approx 1$. As a consequence, we have 

\begin{eqnarray}
B_{\rm seed}&\approx&\int_{t_i}^{t} \dd \tau \frac{e}{\kappa}\left(n \delta\Omega_{ep}+\Omega \delta n _{ep}\right)\\\label{2integrals}
 &=& \int_{t_i}^{t} \dd \tau \frac{e}{\kappa_{\rm Coul}}\left(n \delta\Omega_{ep}+\Omega \delta n _{ep}\right)+\int_{t_i}^{t} \dd \tau \frac{e}{\kappa_{\rm Th}}\left(n \delta\Omega_{ep}+\Omega \delta n _{ep}\right) \\
 &\approx& \int_{t_i}^{t} \dd \tau \frac{e}{\kappa_{\rm Coul}}\left(n \delta\Omega_{ep}+\Omega \delta n _{ep}\right)
\end{eqnarray}
with $T(t_i)=T_i$, $T(t_1)=T_1$ and $T(t_2)=T_2$.  For the temperature
range considered, $T<100~{\rm eV}$, we have the following rough
estimate of the seed magnetic field generated by Coulomb conductivity

\begin{eqnarray}
\dd B_{\rm seed}&\approx&1.7\times{\left(k_B T\right)}^2
\frac{\beta\ln(\Lambda)}{\sqrt{\epsilon_0 \hbar^3 c^5}}{
\left(\frac{m_e c^2}{k_B T}\right)}^{1/2}
{\left(\frac{\delta T}{T}\right)}^2 \frac{\ell_\gamma^3}{\lambda^4}\frac{\delta n_{ep}}{n}c\dd t.
\end{eqnarray}

\section{Strength and coherence length of $B_{\rm seed}$ from primordial fluctuations}\label{appendix_fluctuations}

In this Appendix, for a given physical quantity $M$, $M_i$ refers to the
current value of the quantity at Hubble radius, $M$ the value at the time of the generation
of the magnetic field, and $M_0$ its current value. 

\subsection{Primordial fluctuations}

Let us consider a fluctuation entering the Hubble radius at radiation domination far
from 1~eV. The Hubble parameter is \cite{peter2013primordial,berezhiani2004generation}

\begin{eqnarray}\label{Hi}
H_i^{-1}\approx\frac{27~{\rm kpc}}{c}\left(\frac{1~{\rm eV}}{T_i}\right)^2.
\end{eqnarray}

For $T>20$ eV, $1/\ell_\gamma=\sigma_{\rm Th}n_e$, hence,

\begin{eqnarray}\label{lgi}
\ell_{\gamma,i}\approx30~{\rm pc}\left(\frac{1~{\rm eV}}{T_i}\right)^3.
\end{eqnarray} 

In the radiation dominated era, $a\propto t^{1/2}$. As a consequence,
$H_i^{-1}=2t_i$ and we deduce from Eqs.~(\ref{Hi}) and (\ref{lgi})
that

\begin{eqnarray}\label{lgi_bis}
\frac{\ell_{\gamma,i}}{ct_i}\approx2.2\times 10^{-3}\times\frac{1~{\rm eV}}{T_i}.
\end{eqnarray} 

By definition, the wavelength of the fluctuation entering the Hubble radius
is given by $\lambda_i=cH_i^{-1}$. As we are interested in
$\lambda_0$, which gives us the wave-length of the fluctuation today
and thus the correlation-length of the magnetic field today, let us
express the temperature $T_i$ as a function of
$\lambda_M=\lambda_0/1~{\rm Mpc}$. We know that the temperature
evolves as $T\propto\lambda^{-1}$, so

\begin{eqnarray}\label{TiTf}
\frac{T_i}{T}=\frac{\lambda}{\lambda_i}
\end{eqnarray}

and $\frac{T_i}{T^{(0)}_\gamma}=\frac{\lambda_0}{\lambda_i}$ where
$T^{(0)}_\gamma=2.7255~{\rm K}$ is the temperature of the CMB today
(see \cite{ade2013planck} for the latest evaluation). We can deduce
that

\begin{eqnarray}\label{ti110}
T_i=110\times\lambda^{-1}_M~{\rm eV}.
\end{eqnarray}

\subsection{Evolution of the fluctuations}

For the temperature range considered, the seed magnetic fields are
generated during radiation dominated era. Now, let us focus at the
time in which the seed magnetic fields are generated. As a consequence,
as the temperature evolves as $T\propto a^{-1}$,

\begin{eqnarray}\label{titf}
\frac{t}{t_i}=\left(\frac{T_i}{T}\right)^2. 
\end{eqnarray}

Using the fact that $\ell_\gamma\propto T^{-3}$, we have

\begin{eqnarray}\label{gfgi}
\frac{\ell_\gamma}{\ell_{\gamma,i}}=\left(\frac{T_i}{T}\right)^3. 
\end{eqnarray}
In order to keep our fluctuation unaffected by Silk damping, and to have a vorticity strong enough to generate our seed magnetic field, we take
$\lambda\approx\ell_d$ where $\ell_d$ is the diffusion length of the
photon. This diffusion length is equal to
\cite{peter2013primordial,berezhiani2004generation}

\begin{eqnarray}\label{bilan_1}
\ell^2_d&=&\frac{2}{3}ct \ell_{\gamma}
\end{eqnarray}

From  Eqs.~(\ref{titf}), (\ref{gfgi}) and (\ref{bilan_1}), we have

\begin{eqnarray}\label{ldi}
\ell^2_d=\frac{2}{3}ct_i \ell_{\gamma,i} \left(\frac{T_i}{T}\right)^5,
\end{eqnarray}

from Eqs.~(\ref{TiTf}) and (\ref{ldi}), and by remembering that
$\lambda_i = H_i^{-1}=2ct_i$, we get

\begin{eqnarray}\label{6cti}
\left(\frac{T}{T_i}\right)^3=\frac{\ell_{\gamma,i}}{6ct_i}.
\end{eqnarray}

\subsection{Useful expressions}

From Eqs.~(\ref{lgi_bis}), (\ref{ti110}) and (\ref{6cti}), we can deduce that

\begin{eqnarray}\label{tti}
\frac{T}{T_i}&=&0.015\times\lambda^{1/3}_M,\nonumber\\
            \frac{T}{1~{\rm eV}}&=&1.65\lambda^{-2/3}_M,
\end{eqnarray}

and from Eqs.~(\ref{lgi_bis}) and (\ref{ti110}), that
\begin{eqnarray}\label{ptt}
\frac{\ell_{\gamma,i}}{ct_i} = 2\times 10^{-5}\lambda_M.
\end{eqnarray}

We finally get three last formula that are used in
Section~(\ref{resulting_B}) to evaluate the strength and the
coherence-length of the magnetic field seed. From Eq.~(\ref{tti}), we
get

$$\boxed{\left(\frac{1~{\rm eV}}{T}\right)^{1/2}=0.85\times\lambda^{1/3}_M}$$

from Eq.~(\ref{bilan_1}), we have

$$\boxed{\frac{ct\ell_\gamma}{\lambda^2}=\frac{3}{2}}$$

and using Eqs.~(\ref{gfgi}), (\ref{ptt}) and (\ref{tti}), we obtain

$$\boxed{\left(\frac{\ell_\gamma}{\lambda}\right)^2=2\times10^{-3}\lambda^{2/3}_M}$$

\subsection{Resulting seed fields}\label{app_Bfield}

For the temperature range considered, $T<100~{\rm eV}$, we have the following rough estimate of the seed magnetic field

\begin{eqnarray}\label{basis}
\dd B_{\rm seed}&\approx&1.7\times{\left(k_B T\right)}^2
\frac{\beta\ln(\Lambda)}{\sqrt{\epsilon_0 \hbar^3 c^5}}{
\left(\frac{m_e c^2}{k_B T}\right)}^{1/2}{\left(
\frac{\delta T}{T}\right)}^2 \frac{\ell_\gamma^3}{\lambda^4}\frac{\delta n_{ep}}{n} c\dd t\\
\frac{\dd B_{\rm seed}}{1~{\rm G}}&\approx&3.3\times10^{-13}
\frac{\ell^3_\gamma}{\lambda^4}T^{3/2}\frac{\delta n_{ep}}{n}c\dd t
\end{eqnarray}

We can estimate $\frac{\delta n_{ep}}{n}$ with \cite{takahashi2008electromagnetic},

\begin{eqnarray}
\delta n_{ep}&\approx& \frac{\sigma_{Th}\epsilon_0}{e^2 c}\vec{\nabla}.\rho_\gamma\left(\delta v_{\gamma b}-\frac{1}{4}\vec{v}_b .\Pi_\gamma\right)
\end{eqnarray}

Let us neglect the anisotropic pressure term $\Pi_{\gamma}$, we are reduced to

\begin{eqnarray}
\delta n_{ep}&\approx& \frac{\sigma_{Th}\epsilon_0}{e^2 c}\vec{\nabla}.\left[\rho_\gamma\delta v_{\gamma b}\right].
\end{eqnarray}

With the help of \cite{takahashi2008electromagnetic}, we can express $v_{\gamma b}$,

\begin{eqnarray}
\delta v_{\gamma b} &\approx& \frac{\pi m_p c^2}{4\alpha\hbar c}\frac{\delta_\gamma}{\lambda^2}.
\end{eqnarray}

As a consequence, we get the final expression. 

\begin{eqnarray}\label{delta_n}
\frac{\delta n_{ep}}{n}&=&\frac{136}{\lambda^2 T^3}
\end{eqnarray}

Then, substituting Eqs.~(\ref{TiTf}), (\ref{titf}) and (\ref{delta_n}) in Eq.~(\ref{basis}), integrating between $T_i$ and $T$ and knowing that $\frac{T}{T_i}\ll1$, we finally have

$$\boxed{\frac{B_{\rm seed}}{1~{\rm G}}\approx3.7\times10^{-52}\times T^{5/2}}$$

\section{Typical time scale of diffusion}\label{diffusion}

Let us now estimate the typical time of diffusion $\tau$ in Eq.~(\ref{MHD}) to check if the process we are studying is sustainable.

\begin{eqnarray}
\tau^{(\rm Coul)}&=&\mu_0\kappa_{\rm Coul}\times\left(\frac{\lambda}{2\pi}\right)^2\\
    &=&\frac{6\pi\sqrt{2\pi}}{\alpha \ln(\Lambda)}\left(\frac{k_B T}{m_e c^2}\right)^{1/2}\left(\frac{k_B T}{\hbar c}\right)\frac{1}{c}\left(\frac{\lambda}{2\pi}\right)^2\\
    &=&4.3\times10^{-4}\times T^{3/2}_{\rm eV}\times\lambda^2\\
    &=&5.8\times10^{34}\times\frac{1}{T^{7/2}_{\rm eV}}
\end{eqnarray}

As a consequence, $\tau^{(\rm Coul)}_{\rm T=100~eV}=5.8\times10^{27}$ s. As the age of the Universe is approximately $4.36\times10^{17}$ s, we do not need to worry about the dissipation of our seed magnetic fields for $T<72$~keV.

\section{Amplification of the seed fields}\label{amplification}

First, let us consider a seed magnetic field $B_{\rm seed}$ generated
in the temperature range considered earlier just before
recombination. The conductivity of the primordial plasma after
recombination is very high, therefore the flux of the magnetic field
is conserved and the magnetic field evolves as $B\propto a^{-2}$
\cite{martin2008generation}. This property is taken into consideration
until $z\approx 10$. This redshift corresponds to the lowest redshift
at which the collapse of the early structures, such as the first
proto-galaxies, is expected to happen
\cite{miralda2003dark,naoz2006first,ricotti2008fate,gao2007first}. As
a consequence, the magnetic field decreases by 2 orders of magnitude
until $z\approx 10$.

Then, when the first structures collapse, the seed magnetic field
gains 4 orders of magnitude using a pre-amplification due to
adiabatic compression in the pre-galactic medium before galactic
formation \cite{lesch1994protogalactic}. The combination of the two
effects described above in this Section amplifies the seed magnetic
field generated before recombination by 2 orders of magnitude.

Finally, for $z\in[0,10]$, large-scale magnetic fields are maintained
in galaxies and clusters by a dynamo amplification. For spiral
galaxies, the most popular galactic dynamo model is the
$\alpha\omega$-dynamo whereas other theories are used for elliptical
galaxies and clusters (see \cite{widrow2002origin,kulsrud1999critical}
for a detailed and critical discussion on the different dynamos). The
dynamo amplification $\mathcal{A}$, from an initial field strength
$B_1$ at time $t_1$, corresponding to the end of galaxy formation, to
a galactic field strength $B_0$ at time $t_0$, \textit{i.e.} our
current epoch, is given by
\begin{eqnarray}
\mathcal{A}=\frac{B_0}{B_1}=e^{\Gamma (t_0-t_1)},
\end{eqnarray} 
where $\Gamma$ is the amplification rate. $\Gamma$ is highly dependent
on the cosmological model chosen. However, it is usually found in the
literature that $0.2~{\rm Gyrs}<\Gamma^{-1}<0.8~{\rm Gyrs}$
\cite{widrow2002origin}. Small values of $\Gamma^{-1}$, such as
$\Gamma^{-1}=0.2~{\rm Gyrs}$, are related to processes on small
scales, \textit{i.e.} at intra-galactic scales, and involve fast
amplifications. Large values of $\Gamma^{-1}$, for example
$\Gamma^{-1}=0.8~{\rm Gyrs}$, are related to processes on large
scales, at inter-galactic scales, and trigger slow amplifications. The
galactic dynamo amplification can thus lie in the interval
$[10^{7},10^{28}]$. As a consequence, magnetic fields on large scales
are going to be less amplified than those on small scales. This is
coherent with the data, which reveal that there not much difference of
strength between this two types of magnetic fields. This also
guarantee that the magnetic fields in the
inter-galactic medium are not to strong compared to data.

\end{document}